# Atomically Sharp, Closed Bilayer Phosphorene Edges by Self-Passivation

Sol Lee,[1,2,†] Yangjin Lee,[1,2,†,*] Li Ping Ding,[3,4,†] Kihyun Lee,[1] Feng Ding,[3,5,*] and Kwanpyo Kim[1,2,*]

[1] Department of Physics, Yonsei University, Seoul 03722, South Korea

[2] Center for Nanomedicine, Institute for Basic Science, Seoul 03722, South Korea

[3] Center for Multidimensional Carbon Materials, Institute for Basic Science, Ulsan 44919, South Korea

[4] Department of Optoelectronic Science & Technology, School of Electronic Information and Artificial Intelligence, Shaanxi University of Science & Technology, Xi'an 710021, China

[5] Department of Materials Science and Engineering, Ulsan National Institute of Science and Technology, Ulsan 44919, South Korea

† These authors contributed equally to this work.

*Address correspondence to Yangjin Lee (yangjinlee@yonsei.ac.kr), Feng Ding (f.ding@unist.ac.kr), and Kwanpyo Kim (kpkim@yonsei.ac.kr)

ABSTRACT

Two-dimensional (2D) crystals' edge structures not only influence their overall properties but also dictate their formation due to edge-mediated synthesis and etching processes. Edges must be carefully examined because they often display complex, unexpected features at the atomic scale, such as reconstruction, functionalization, and uncontrolled contamination. Here, we examine atomic-scale edge structures and uncover reconstruction behavior in bilayer phosphorene. We use in situ transmission electron microscopy (TEM) of phosphorene/graphene specimens at elevated temperatures to minimize surface contamination and reduce e-beam damage, allowing us to observe intrinsic edge configurations. Bilayer zigzag (ZZ) edge was found the most stable edge configuration under e-beam irradiation. Through first-principles calculations and TEM image analysis under various tilting and defocus conditions, we find that bilayer ZZ edges undergo edge reconstruction and so acquire closed, self-passivated edge configurations. The extremely low formation energy of the closed bilayer ZZ edge and its high stability against e-beam irradiation are confirmed by first-principles calculations. Moreover, we fabricate bilayer phosphorene nanoribbons with atomically-sharp closed ZZ edges. The identified bilayer ZZ edges will aid in the fundamental understanding of the synthesis, degradation, reconstruction, and applications of phosphorene and related structures.

INTRODUCTION

Edges play pivotal roles in the formation of two-dimensional (2D) crystals. Synthesis and etching processes in 2D crystals mainly take place at exposed edges and the successful engineering of edge structures can lead to the controllable synthesis of high-quality 2D crystals.[1-6] Moreover, dangling bonds at edges' termination geometries cause them to have different electrical and chemical properties from those of the crystals' basal planes, which can be exploited for many applications.[7-10] To investigate the edges of 2D crystals at atomic resolution, transmission electron microscopy (TEM) has been previously utilized.[1-3, 11-19] Atomic-resolution TEM imaging showed predominant zigzag (ZZ) or armchair edge configurations, atomic-scale vacancies and reconstructions, and edge functionalization in graphene, hexagonal boron nitride (h-BN), and transition metal dichalcogenides.[1-3, 11-19] Moreover, these studies demonstrated that crystal edges at the atomic scale are often far from the simple cleaving geometries of 2D crystals, so they should be carefully examined.

Phosphorene, a mono-elemental 2D crystal composed of phosphorus, is a promising building block for optoelectronics.[20-23] Phosphorene also possesses anisotropic puckered structures. Its physical properties strongly depend on its crystal lattice direction. Phosphorene exhibits strong anisotropic in-plane charge transport behavior and one-dimensional diffusion on its surface.[21, 24-28] Although ZZ terminations are the predominant configurations in phosphorene,[29-35] its edges have not been observed at the atomic to a significant degree. In particular, it is difficult to image phosphorene edges because they are degraded by ambient exposure and unintended alterations to them during characterization. Moreover, the phosphorus bonds can produce various types of nanostructures.[36-39] Unexpected edge reconstructions may be identified by careful measurement at an atomic resolution.

In this study, we identified an atomic-scale edge reconstruction in bilayer (2L) phosphorene. By employing aberration-corrected TEM with an in-situ heating holder, we successfully fabricated and imaged atomically sharp phosphorene edge configurations. The phosphorene specimens were supported by a graphene membrane, which served as an ideal TEM imaging substrate. We found that ZZ edges in 2L phosphorene were highly stable under e-beam irradiation. Via first-principles calculations and extensive TEM imaging with simulations, we determined that ZZ edges in 2L phosphorene were self-passivated by forming covalent bonds between the two edges. The closed, self-passivated ZZ edges are the most stable edge configuration and are resistant to high-energy e-beam irradiation damaging. We also successfully fabricated 2L phosphorene nanoribbons in situ with predominantly atomically sharp, stable, closed ZZ edges. This study enhances our fundamental understanding of various aspects of phosphorene and related structures, including their synthesis, electronic applications, and chemical processes related to their edges.

RESULTS AND DISCUSSION

Phosphorene/graphene specimens were prepared on in situ TEM microchips (Supp. Figs. S1 and S2). Samples were fabricated by mechanically exfoliating graphene and black phosphorus (BP) flakes. More detailed information about how the samples were prepared can be found in the Methods and Supporting Information sections. Graphene membranes served as a chemically inert, ultrathin imaging support for the phosphorene during TEM observation.[31]

We observe the controlled etching of phosphorene at elevated temperatures, which leads to the formation of ultra-thin phosphorene regions (Fig. 1a). We mainly chose 270 °C as the heating temperature because heating phosphorene above 300 °C accelerates phosphorene removal.[40, 41] Most of the etching occurred on the samples' exposed edges and phosphorene at the basal plane were relatively stable during TEM imaging. TEM imaging showed that etching thinned the samples and caused the formation of 2L and monolayer (1L) regions (Fig. 1a). The complete local etching of the phosphorene exposed the graphene membrane (Figs. 1b and 1d), which made it easier to count the number of layers. 1L and 2L phosphorene regions can be identified by intensity patterns (Figs. 1e and 1f).[31]

Etching at and the atomic-scale edge configurations of ZZ and armchair lattice terminations were different because of phosphorene's anisotropic crystal structure and edge stability. Phosphorene had a pronounced anisotropic reaction to etching. The etching rate at the armchair terminations was higher than that at the ZZ terminations because armchair configurations have lower energy barriers for atomic sputtering.[31] The time-dependent edge evolution of the first and second layers in phosphorene showed that the location of exposed edges changed much faster horizontally than vertically (Figs. 1h and 1i, Supp. Movie 1, Supp. Fig. S3). ZZ terminations had more crystalline configurations while armchair edges had mostly amorphous-like edge terminations. The samples' anisotropic responses to etching were more pronounced at elevated temperatures than at room temperature.[31] Conducting TEM imaging at elevated temperatures reduces sample contamination, allowing for better observation of etching responses.[12, 42]

Merged 2L ZZ edges behaved differently than ZZ edges in individual phosphorene layers. During etching, sample edges in individual layers can overlap, resulting in 2L edge termination. Once formed, 2L ZZ edges are stable under e-beam irradiation (Supp. Movie 1). We observed very little change in edge location and morphology over ~100 seconds. The merged ZZ edges (2L ZZ edge) are more stable than unmerged ZZ edge terminations in individual layers (Fig. 1j). Graphene lattices were removed by signal filtering which was achieved by Fourier transform to better observe the phosphorene's structure (Fig. 1k–n). Time-series TEM images (Supp. Fig. S4) showed the formation process of 2L ZZ edges, in which the segment of atomically-sharp ZZ edge readily spreads out to consume the rough edge segment. The occasionally-observed kinks at ZZ edge also showed relatively high stability under e-beam (Supp. Fig. S4). The formation of atomically sharp, straight ZZ configurations is highly reproducible as demonstrated in multiple TEM images in Fig. 1o and Supp. Fig. S5.

The fact that ZZ edges were more stable in 2Ls than 1Ls indicates that the edges of the top and bottom layers interact strongly, such as via covalent bonds. Simulated TEM images of non-reconstructed pristine 2L ZZ terminations did not match observed edge images (Supp. Fig. S6), which supported the existence of interlayer interactions. Therefore, we explored various types of reconstructed ZZ edges in 2L phosphorene using the Crystal structure AnaLYsis by Particle Swarm Optimization (CALYPSO) global search software package[43, 44] with first principles calculations. We examined pristine 2L and various types of closed edge configurations (Fig. 2a and Supp. Fig. S7). When exposed-edge atoms in the top and bottom layers connect to form closed edges, the local strain near the edge occurs to accommodate the formation of covalent bonds. Because of the edge-associated strain effect, the local stacking configuration near the edge may shift from energetically-favored AB stacking configuration.[23, 45] The edges were classified as either ZZ-AC or ZZ-AA according to whether the layers shifted perpendicularly or in parallel relative to the edge termination, respectively. We named closed edge configurations as either ZZ-AC# or ZZ-AA#, where # is an assigned identifying number. ZZ-AC configurations had edges with a near-edge stacking order similar to AC stacking in which two layers are shifted perpendicularly relative to the edge termination. ZZ-AA configurations had edges with a near-edge stacking order similar to AA stacking in which two layers are shifted in parallel to the edge termination (Supp. Fig. S8). We observed that the stacking order gradually transitioned from AB-type stacking to another configuration near atomically sharp ZZ edges, but not near rough and therefore open ZZ edges (Supp. Fig. S9), which further supports the conclusion that closed-edge formation caused local strain.

Fig. 2b shows the calculated edge formation energies of various 2L ZZ edge structures. The edge formation energy of a specific structure is defined as:

$$E_f = \frac{2E_2 - E_1}{2L} \quad (1)$$

where $E_1$ is the energy of a double-sized phosphorene nanoribbon (PNR) with two H-terminated edges, $E_2$ is the energy of a half-width edge-reconstructed PNR with one H-terminated edge, $L$ is the length of a 2L PNR edge, and the factor of 2 accounts for the fact that two identical edges are produced after cutting a large PNR. ZZ-AC1 is the most stable one in all these configurations, including the open 2L ZZ edge. In the ZZ-AC1, ZZ-AC2, and ZZ-AA1 closed-edge reconstructions, every phosphorus atom possessed the three nearest neighbor atoms as that in black phosphorus bulk and the bonding angles are also close to that in black phosphorus bulk. So, these closed edge configurations have no unsaturated covalent bond, resulting in relatively stable configurations (Fig. 2b). On the other hand, some edge atoms in ZZ-AA2 and ZZ-AA3 edges possessed two nearest neighboring atoms (Supp. Fig. S7), leading to larger edge formation energies. ZZ-AC1 edges were formed when edge atoms in one layer were drawn toward those in another layer, forming interlayer phosphorus bonds (Fig. 2c). Such a simple process suggests that the close 2L edges can be formed in 2L phosphorene naturally or by external manipulation.

We performed extensive atomic-resolution TEM imaging and simulations to confirm the formation of closed 2L edge reconstruction. To precisely identify the edges' three-dimensional structures at an atomic resolution, we performed TEM imaging of edges at various defocus values and tilting configurations. Fig. 3a and Supp. Fig. S10 show experimental and simulated TEM images of various types of ZZ configurations. Experimental images agreed well with simulated images of ZZ-AC1 closed ZZ edges but were not consistent with those for other types of edges. We also performed TEM imaging under various tilting conditions, which further corroborated our results. Experimental and simulated TEM images of ZZ-AC1 edges were well matched (Fig. 3b and Supp. Fig. S11).

Phosphorus forms various polymorphs and allotropes due to the versatile nature of phosphorus bond.[36, 37, 46] Simple edge terminations that were not reconstructed in phosphorene had unsaturated bonds and were thermodynamically unstable than passivated edges (Fig. 2b). In contrast to the previous scanning tunneling microscopy and TEM imaging results on reconstruction behavior associated with individual layers,[29, 30, 33] our study experimentally confirms interlayer-associated edge reconstruction in 2L phosphorene. We note that a similar type of edge reconstructions has previously been observed in 2L graphene and h-BN.[2, 13-15]

The stability of 2L ZZ edge under e-beam irradiation can be also explained by their closed edge reconstruction. We first consider the knock-on atomic sputtering by e-beam irradiation. We calculated the energy barriers of atomic sputtering process using the climbing images-nudged elastic method.[47-49] Without closed edge formations, the energy barrier for removing phosphorus atoms from terminated open ZZ edges was determined to be 5.59 eV (Figs. 4a and 4b). On the other hand, the energy barrier for atomic sputtering ZZ-AC1 edges was determined to be 6.31 eV. Then, we calculated the differential cross-section of energy transfer events as a function of transferring energy to phosphorus atoms under 80 kV e-beam irradiation (Fig. 4c).[31, 50] The detailed cross-section calculation methods can be found in Supp. Note S1. We found that the maximum transferred energy to phosphorus atoms by 80-kV incident e-beams was $T_{max}$ = 6.11 eV from the static model or approximately $\widetilde{T}_{max}$ = 6.5 eV from the dynamical lattice model, as shown in Fig. 4b and 4c. The maximum transferred energy is bigger than the knock-on energy barrier for pristine ZZ edge and very close to that for ZZ-AC1 edge. Moreover, the calculated knock-on cross-section for atoms at the ZZ-AC1 edge is five times smaller than that for atoms at the pristine ZZ edge under the dynamical lattice model as shown in Fig. 4d. Under the electron dose rate we employed for TEM imaging (approximately 2.0 × $10^6$ e$^-$ nm$^{-2}$ s$^{-1}$), ZZ-AC1 is found to be stable for more than 100s as shown in Supp. Fig. 12. This result is consistent with our observation. We note that an energy barrier for atomic sputtering process obtained from molecular dynamics is considered to be more accurate and the utilized nudge elastic band method usually underestimates an energy barrier.[51, 52] If we consider the possible underestimation for the calculated energy barriers, the expected stability of the ZZ-AC1 edge against knock-on damage is greater than our current analysis.

Chemical etching process, which is driven by chemical reaction with surface residues or residual gases in TEM column,[12, 53, 54] also needs to be considered as another etching mechanism. Although we tried to minimize the effect of chemical etching by utilizing the graphene support and imaging at elevated temperatures[12], the chemical etching appears to play an important role in our study, especially during the initial etching of BP samples. The closed 2L ZZ edge configuration can also provide the significant protection against the chemical etching because saturated covalent bonds at the closed edge hinder the adsorption of reactive chemical species. On the other hand, the exposed edges suffer from chemically active dangling bonds and are more prone to the chemical etching process.

Atomically well-defined phosphorene nanostructures can be fabricated using closed 2L ZZ edges. For example, we successfully fabricated bilayer phosphorene nanoribbons (BPNRs) by e-beam-based nanofabrication (Fig. 5 and Supp. Movie 2). Fig. 5a shows a series of HR-TEM images showing the formation of BPNRs with closed ZZ edges. Etching usually starts in from regions with surface residues and prolonged e-beam irradiation forms nanoribbons with atomically sharp closed 2L ZZ edges (Fig. 5b). We believe that better controlling nucleation during etching will increase the degree to which nanoribbon width and density can be controlled.

The electrical properties of BPNRs were investigated by first-principles calculations. We performed the calculations for both BPNRs with ZZ-AC1 edges and with unreconstructed pristine ZZ edges (Figs. 5c and 5d, respectively). The electronic band structures and densities of states were calculated based on the Heyd-Scuseria-Ernzerhof hybrid exchange-correlation function,[55, 56] which is reliable for electrical property calculations.[25, 57, 58] We found that the electronic band structure of BPNRs strongly depended on their edge configurations. 2L phosphorene is a semiconductor with a direct bandgap of approximately 1.1 eV.[23] Without edge reconstructions, BPNRs exhibit metallic behaviors with in-gap states associated with edges (Fig. 5c). However, BPNRs with closed edge configurations have some of the semiconducting nature of phosphorene because their edge states are inactivated (Fig. 5d). This result demonstrates how BPNRs' edge configurations affect their electronic and chemical applications.[58-61] Theory related to BPNRs with closed edge configurations has recently been reported.[62, 63]

CONCLUSIONS

We examined atomic-scale edge reconstructions in 2L phosphorene. Atomic resolution TEM imaging showed that closed, self-passivated ZZ edges formed in 2L phosphorene. These edges were the most stable configuration of various tested types and were robust to high-energy e-beam irradiation. This result was confirmed by first-principles calculations. Moreover, we successfully fabricated 2L phosphorene nanoribbons in situ with predominantly atomically sharp, stable, closed ZZ edges. These identified stable, closed ZZ edges will help increase the fundamental understanding of the synthesis, degradation, reconstruction, and applications of phosphorene.

METHODS

**Sample Preparation:** Phosphorene/graphene vertical heterostructures were prepared via dry transfer inside an inert atmosphere glove box in which oxygen and water vapor concentrations were maintained below 0.1 ppm. BP flakes (Smart Elements) and graphite were mechanically exfoliated on semitransparent polydimethylsiloxane supports. The flakes' local thicknesses were identified using a DM-750M optical microscope (Leica) in transmission mode. Graphene flakes with less than 5 layers were transferred to heating E-chips (Protochips, Inc.). The samples were then mounted on in-situ Fusion heating holders (Protochips, Inc.) that were heated to 500°C for 3 hrs in a vacuum of $\sim 10^{-7}$ Torr to remove residues from the graphene flakes. The identified phosphorene flakes were transferred onto a pre-annealed graphene flake sample to fabricate phosphorene/graphene heterostructures. For some TEM samples, $MoS_2$ flakes were pre-transferred onto E-chips before being transferred onto graphite to reduce the surface roughness of and enhance the adhesion between graphite and phosphorene flakes.[64]

**TEM Imaging and Simulations:** TEM images were acquired using a JEOL-ARM 200F TEM microscope operated at 80 kV that was equipped with a double Cs-aberration corrector. The vacuum level was approximately $1 \times 10^{-7}$ mbar at the specimen column. We used a Protochip Fusion in situ holder for observation of samples at elevated temperatures. Samples were heated at a rate of 1°C/s and maintained at 270°C during TEM imaging. Most TEM time series images were captured with an exposure time of 0.25 s under an electron dose rate of approximately $2.0 \times 10^6$ $e^-$ $nm^{-2}$ $s^{-1}$. TEM image simulations were performed using MacTempas and abTEM software.[65] $C_s$ = -10μm, convergence angle = 0.10mrad, objective aperture radius = 1.5Å$^{-1}$, and mechanical

vibration = 0.5Å (or comparable simulation conditions) under various defocus values and sample tilting configurations were used for TEM image simulations.

**Calculations:** Phosphorene ZZ edge reconstructions were examined by an unbiased global search method and CALYPSO combined with first-principal calculations,[43, 44] which has been successfully used to predict the surface reconstruction of bulk materials[66, 67] and the edge reconstructions of 2D materials.[68, 69] During the global structural search, 60% of the 40 candidates in each generation were generated by a particle swarm optimization algorithm and the other 40% were generated randomly. For simplicity, we assumed that the edges were initially passivated by hydrogen atoms. All of the structures were optimized by density functional theory calculations as implemented in the Vienna Ab initio Simulation Package.[70, 71] The Perdew-Burke-Ernzerhof generalized gradient approximation[72] was used for the exchange-correlation interactions. The projected augmented wave method[71] was used to treat the valence electron-ion core interactions. The density functional theory-D2 method[43] was used to describe the Van de Waals interactions between 2L of phosphorene. We used a plane-wave basis set with a cutoff energy of 450 eV in the calculations. All structures were fully relaxed and the convergence criteria for energy and force were set at $10^{-5}$ eV and $10^{-2}$ eV/Å, respectively. The energy barriers for atomic sputtering of ZZ-AC1 edges were obtained by using the climbing images-nudged elastic method.[47-49] The forces on the images were relaxed until they reached the 0.02 eV/Å threshold.

ASSOCIATED CONTENT

**Supporting Information**.

The Supporting Information is available free of charge at https://pubs.acs.org/doi/....

Detailed calculation methods for knock-on cross-section under static and dynamic lattice models, schematics of the sample fabrication process, optical images of samples, and extra TEM characterizations and analysis. (PDF)

Supporting Movie S1: Formation of closed bilayer phosphorene edges (AVI)

Supporting Movie S2: Formation of BPNRs with closed ZZ edges (AVI)

## AUTHOR INFORMATION

**Corresponding Author**

*Address correspondence to Yangjin Lee (yangjinlee@yonsei.ac.kr), Feng Ding (f.ding@unist.ac.kr), and Kwanpyo Kim (kpkim@yonsei.ac.kr)

**Author Contributions**

[†]S.L., Y.L., and L.P.D. contributed equally to this work.

**Author Contributions**

Y.L., and K.K designed the experiments. S.L. and Y.L. performed experiments and analyzed the data. L.P.D. and F.D. performed calculations. K.L. performed part of TEM image simulations. All of the authors helped write this manuscript.

**Notes**

The authors declare no competing interests.

## ACKNOWLEDGMENT

This work was mainly supported by the Basic Science Research Program at the National Research Foundation of Korea (NRF-2017R1A5A1014862, NRF-2021R1C1C2006785, NRF-2022R1A2C4002559), the 2022 Yonsei Signature Research Cluster Program (2022-22-0004), the Institute for Basic Science (IBS-R019-D1 and IBS-R026-D1), and the Youth Talent Invitation Scheme of the Shaanxi Association for Science and Technology (No. 20190506).

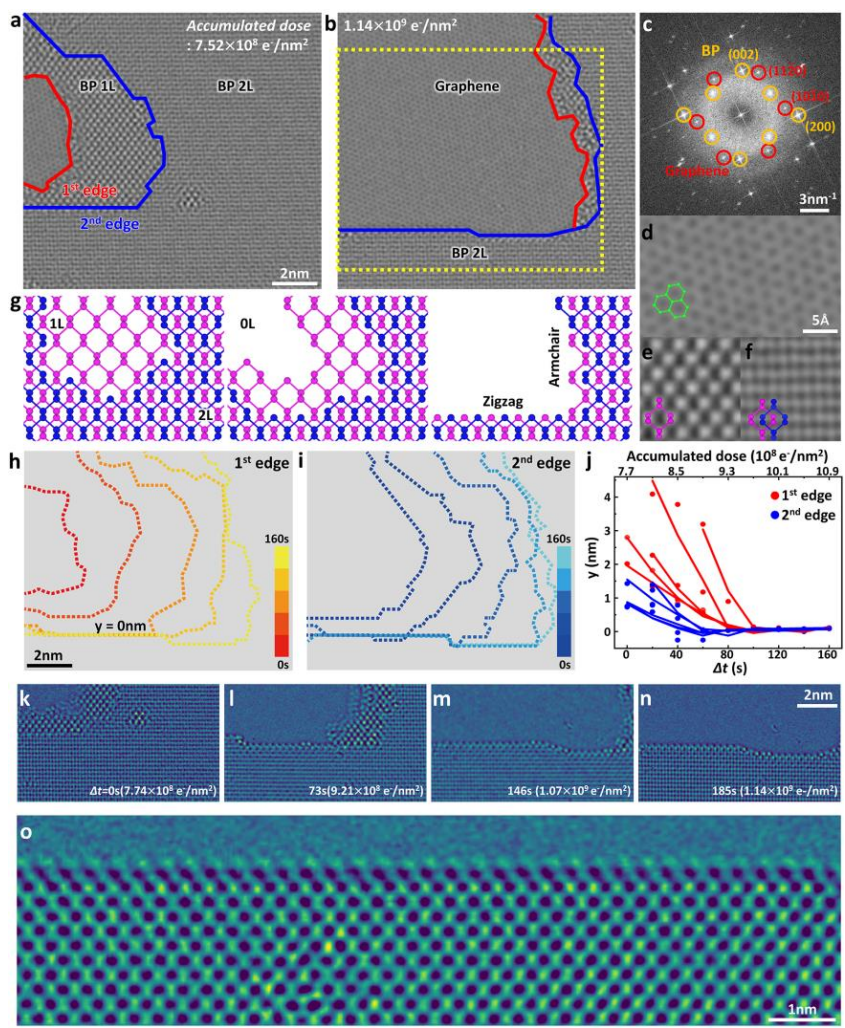

**Figure 1. Formation of atomically sharp 2L phosphorene ZZ edge.** (a) HR-TEM image of 2L phosphorene/graphene The edge terminations of the first and second layers in phosphorene are marked with red and blue dashed lines. (b) HR-TEM image of the same area after prolonged e-beam irradiation The yellow dashed region is the area for analysis in panel h-j. (c) Fast-Fourier-transformed TEM image. The phosphorene diffraction signals are marked with yellow circles. (d) Magnified TEM image of an exposed graphene membrane. (e) HR-TEM image from monolayer phosphorene. (f) HR-TEM image of 2L phosphorene. (g) Schematic of phosphorene etching under e-beam irradiation. Time-dependent edge termination of the (h) first and (i) second layers. (j) Time trajectories of edge locations along the y-axis in panels h and i. TEM images of 2L ZZ edges after (k) $\Delta t = 0$s, (l) 73s, (m) 146s, and (n) 185s. (o) Magnified HR-TEM image of an atomically sharp 2L ZZ edge.

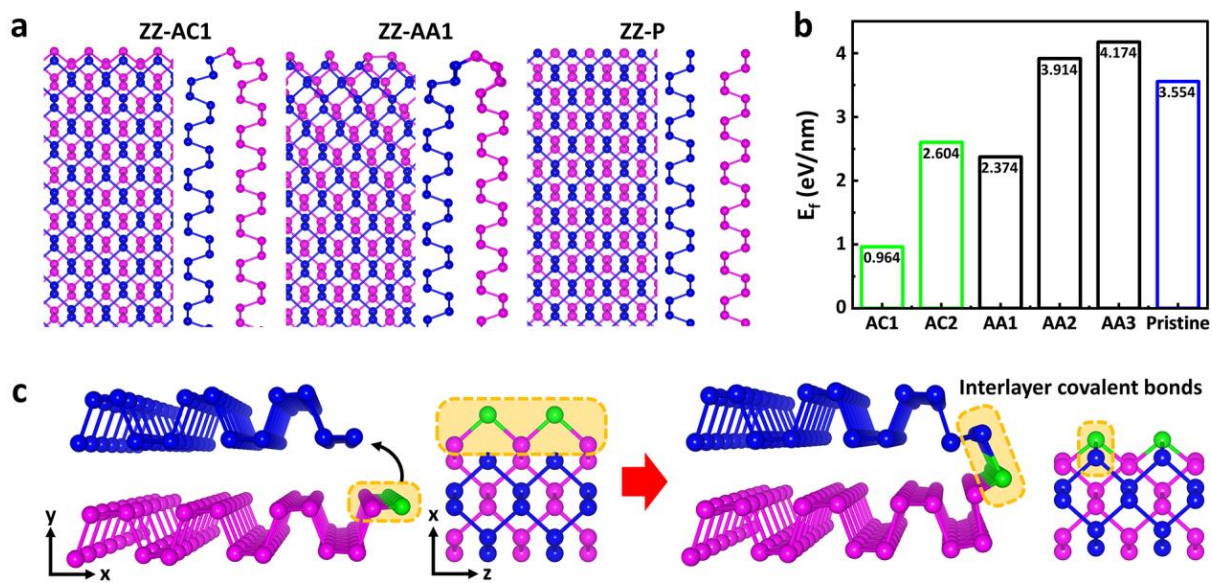

**Figure 2. Theoretical analysis of 2L phosphorene ZZ edge reconstruction.** (a) Top- and side-view schematics of various 2L phosphorene ZZ edges with closed and open configurations. AC and AA indicate the local 2L stacking configuration near the edge. (b) Calculated edge formation energies of various 2L edge configurations. (c) Schematics showing how ZZ-AC1 edges form. Edge atoms (green) in one-layer bend towards another layer and form covalent bonds, resulting in a closed-edge configuration.

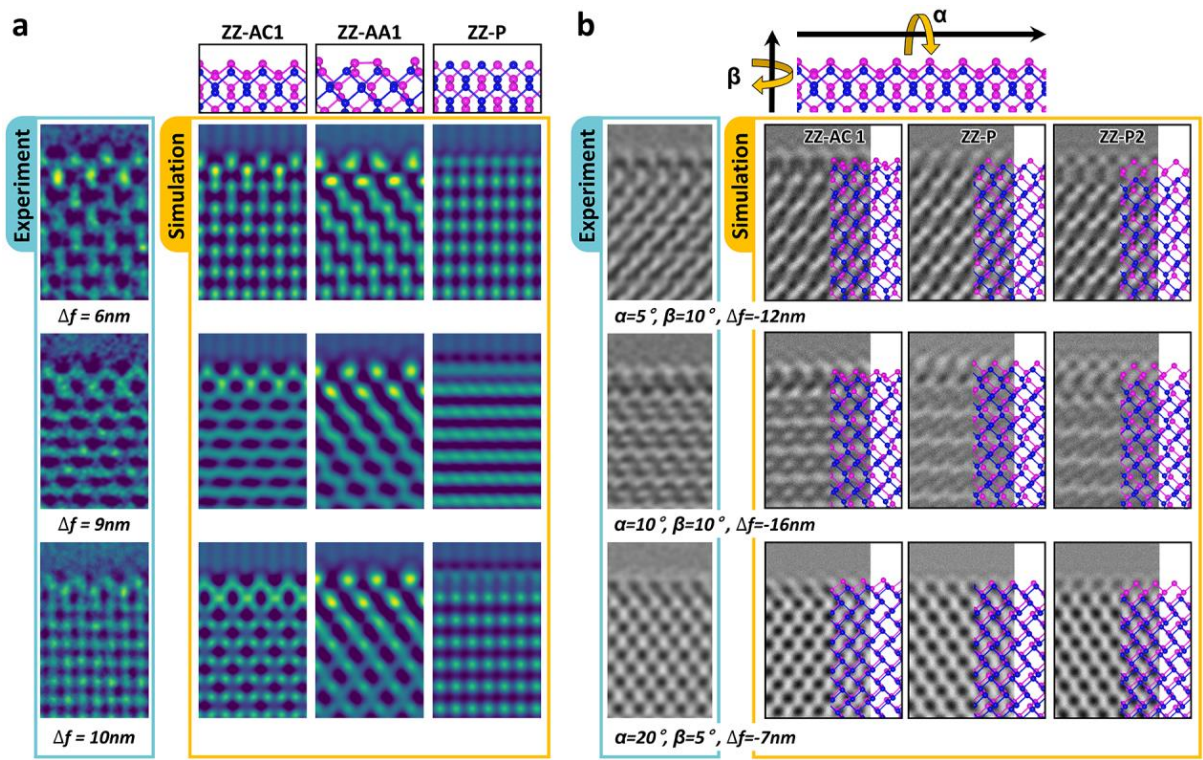

**Figure 3. Experimental confirmation of closed 2L ZZ edge.** (a) Atomic-resolution experimental and simulated TEM images of various types of edge configurations at various defocus values. (b) (Left) Experimental TEM images at various tilting geometry and (right) simulated TEM images of various edge configurations.

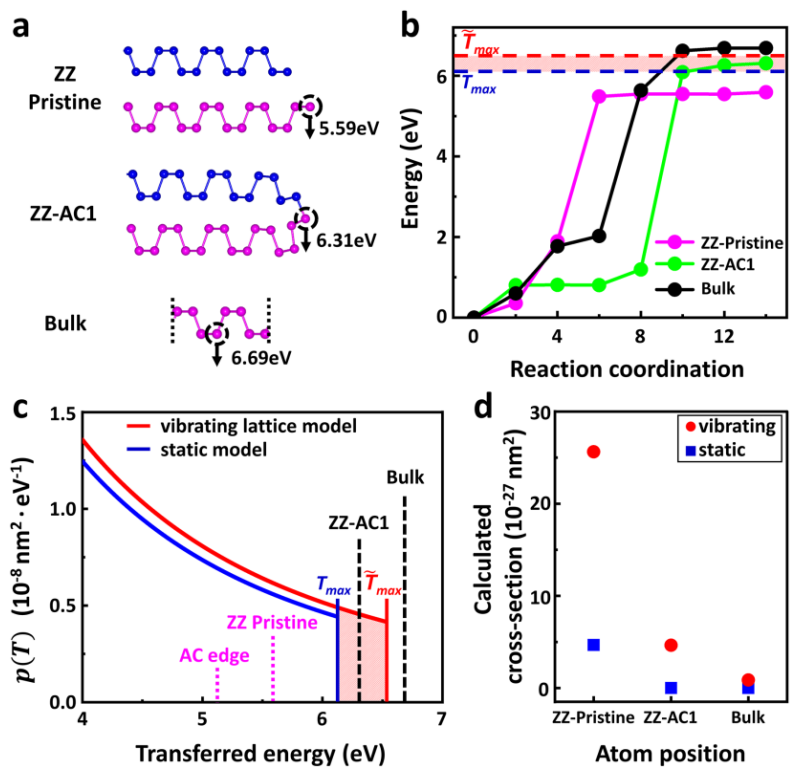

**Figure 4. Stability of closed 2L ZZ edges against e-beam-induced sputtering.** (a) Phosphorus atom removal from pristine edges, closed edges, and bulk regions. The values are the calculated energy barriers for atom removal. (b) Calculated energies at different reaction coordinates during phosphorous atom removal process. The red and blue dashed horizontal lines indicate the values of the maximum energy transferred to phosphorus atoms under e-beam irradiation. (c) Calculated differential knock-on cross-sections as a function of transfer energy under incident 80 keV e-beam irradiation. The blue and red curves are obtained from static and vibrating lattice models, respectively. The values of the maximum energy transfer and energy barriers for atomic sputtering are indicated. (d) Calculated knock-on cross-sections for atoms at ZZ-pristine edge, ZZ-AC1 edge, and bulk under the static and vibrating lattice assumption.

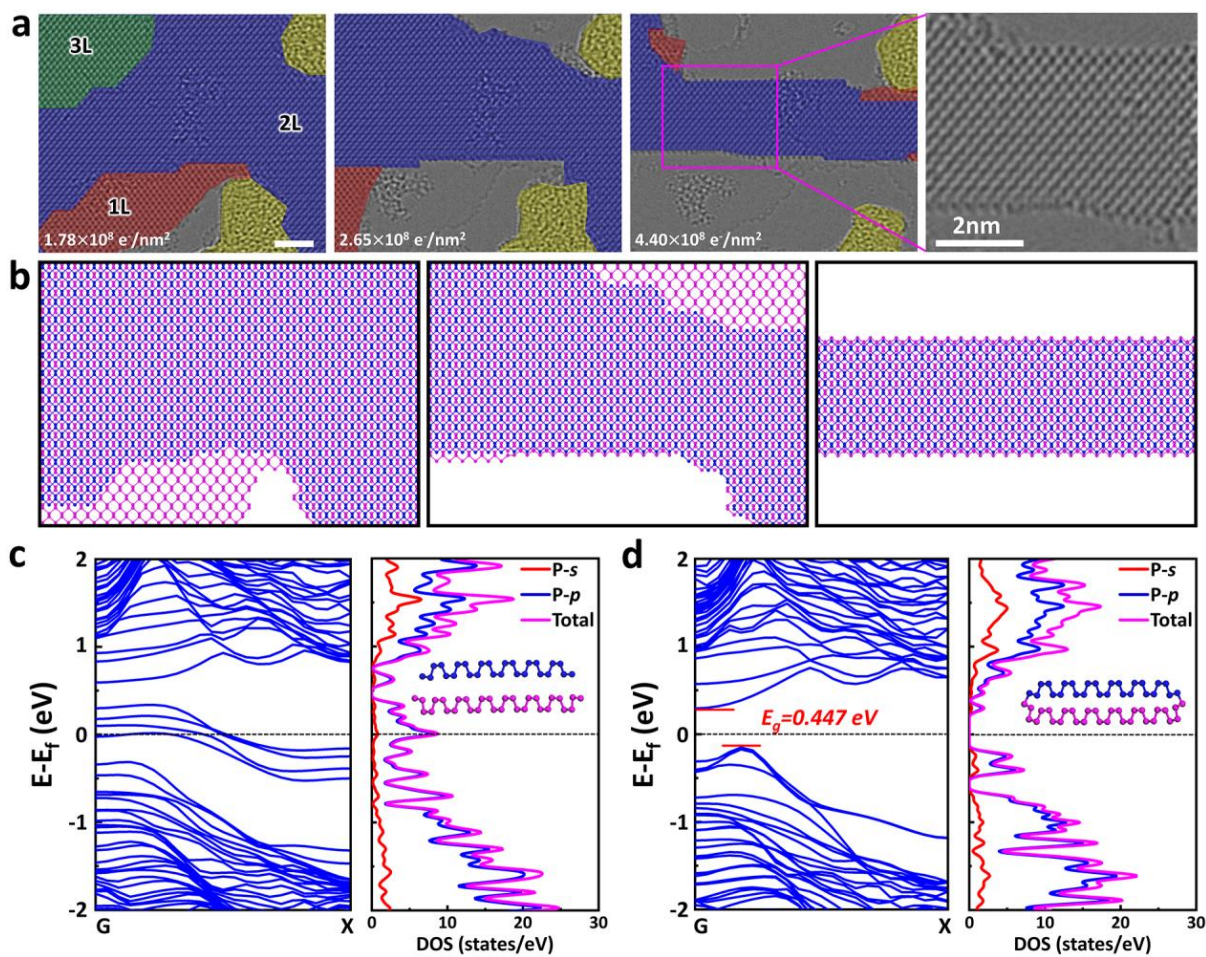

**Figure 5. Formation and properties of 2L phosphorene nanoribbons with closed ZZ edges.** (a) HR-TEM images of 2L phosphorene nanoribbon formation and magnified HR-TEM image of the resulting 2L phosphorene nanoribbon. (b) Schematic of 2L nanoribbon formation by anisotropic etching. (c) Calculated electronic band structure and density of states (DOS) of 2L nanoribbons with pristine open edges. (d) Calculated electronic band structure and density of states of 2L nanoribbons with closed edges.